\documentclass[aps,floatfix,pre,twocolumn]{revtex4-2}
\usepackage[T1]{fontenc}
\usepackage[utf8]{inputenc}
\usepackage{amsmath}
\usepackage{amssymb}
\usepackage{graphicx}
\usepackage{refstyle}
\usepackage{multirow}
\usepackage{float}
\usepackage{soul}
\usepackage[
citecolor=blue,
colorlinks,
linkcolor=blue,
urlcolor=blue,
]{hyperref}
\usepackage[dvipsnames]{xcolor}
\usepackage{dcolumn}
\usepackage{bm}

\begin{document}
\title{Amplitude death in coupled replicator map lattice: averting migration dilemma}	
%
\author{Shubhadeep Sadhukhan}
\email{deep@iitk.ac.in}
\affiliation{
	Department of Physics,
	Indian Institute of Technology Kanpur,
	Uttar Pradesh 208016, India
}

\author{Rohitashwa Chattopadhyay}
\email{crohit@iitk.ac.in}
\affiliation{
  Department of Physics,
  Indian Institute of Technology Kanpur,
  Uttar Pradesh 208016, India
}
\author{Sagar Chakraborty}
\email{sagarc@iitk.ac.in}
\affiliation{
  Department of Physics,
  Indian Institute of Technology Kanpur,
  Uttar Pradesh 208016, India
}

\date{\today}

\begin{abstract}
Populations composed of a collection of subpopulations (demes) with random migration between them are quite common occurrences. The emergence and sustenance of cooperation in such a population is a highly researched topic in the evolutionary game theory. If the individuals in every deme are considered to be either cooperators or defectors, the migration dilemma can be envisaged: The cooperators would not want to migrate to a defector-rich deme as they fear of facing exploitation; but without migration, cooperation can not be established throughout the network of demes. With a view to studying the aforementioned scenario, in this paper, we set up a theoretical model consisting of a coupled map lattice of replicator maps based on two-player--two-strategy games. The replicator map considered is capable of showing a variety of evolutionary outcomes, like convergent (fixed point) outcomes and nonconvergent (periodic and chaotic) outcomes. Furthermore, this coupled network of the replicator maps undergoes the phenomenon of amplitude death leading to non-oscillatory stable synchronized states. We specifically explore the effect of (i) the nature of coupling that models migration between the maps, (ii) the heterogenous demes (in the sense that not all the demes have same game being played by the individuals), (iii) the degree of the network, and (iv) the cost associated with the migration. In the course of investigation, we are intrigued by the effectiveness of the random migration in sustaining a uniform cooperator fraction across a population irrespective of the details of the replicator dynamics and the interaction among the demes.
\end{abstract}
\maketitle

\section{Introduction}
Large spatially distributed populations more often than not form clusters of several subpopulations connected through migration which is one of the important mechanisms in shaping the evolution and bringing forth emergence of cooperation~\cite{1964_Kimura_Wiess_Genetics, 1973_Latter, 2009_Helbing_Yu_PNAS,  2009_WNH_PNAS, 2010YWW, 2011Y, 2011_Roca_Helbing_PNAS, 2012_FN_JSP, 2012WFZW, 2012CSP, 2013BTA, 2018_LELG_NC}. It is easy to envisage that availability of better opportunities~\cite{2013BTA} elsewhere lead individuals to abandon their home and migrate. Moreover, individuals may want to migrate to satisfy their aspirations~\cite{2010YWW,2011_Roca_Helbing_PNAS}. The migration can furthermore depend on the expectations~\cite{2012WFZW} of the individuals. Risk-driven migration~\cite{2012CSP} and success-driven migration~\cite{2009_Helbing_Yu_PNAS, 2011Y} can also promote cooperation effectively. While a random migration arguably weakens the emergence of cooperation by favouring the invasion by defection~\cite{1991_Dugatkin,1993_enquist}, very mildly incentivizing cooperating behaviour~\cite{2021_Sadhukhan_PRR} can overcome this drawback of the random migration. 

Theoretically, emergence and sustenance of cooperation in a collection of subpopulations or demes with active migration between them can be conveniently studied within the paradigm of the evolutionary game theory by using coupled map lattice (CML) models~\cite{1992_Kaneko_Chaos, 2014_Kaneko}. In fact, in such a setting, the phenomena of cooperation, chaos, and synchronization come together to overcome the migration dilemma~\cite{2021_Sadhukhan_PRR}: In a CML of subpopulations of replicators with two actions---cooperate or defect---if all but a few subpopulations have defectors exclusively, the cooperators would not want to migrate lest they should be exploited by the defectors; however, in the absence of any migration, cooperation would not be established across the network of subpopulations and therefore the collective utility gain for the population is denied in the light of cooperators' not risking their relatively higher payoffs. 

The nonlinear dynamics and network dynamics of the evolutionary systems in the context of the interplay between synchronization and cooperation have motivated quite a few recent studies, e.g., the ones on the evolutionary Kuramoto dilemma~\cite{2017_Antonioni_Cardillo_PRL,2018_Liu_Wu_Guan_EPL,2018_Yang_Han-Xin_Zhou} and the one on chaotic agent dynamics~\cite{Rohit_2020_chaos}. In the setting of the CML with chaotic replicator maps, the amplitude variations of the chaotic oscillations of the fraction of the cooperators in the subpopulations are suppressed due to synchronization onto a fixed point of the CML~\cite{2021_Sadhukhan_PRR}. It is natural to draw an analogy with the amplitude death~\cite{SAXENA2012} in coupled oscillators whenever the CML synchronizes onto a fixed point~\cite{2001_JJ_PRE,2003_MM_PRE,2004_LGH_PRE,2005_JAH_PRE,2005_AJH_PRE,2008_PMM_EPJB,2009_LXSXK_Chaos}. Technically, the term amplitude death refers to the situation when the oscillations---either periodic or quasi-periodic or chaotic---of an entire system of coupled oscillators ceases, leading to the stationarity~\cite{Turing1952, Prigogine1968, BarEli1984}. The amplitude death occurs in a wide variety of systems, whether interacting systems are identical~\cite{Reddy1998, Karnatak2010, Karnatak2007, Atay2003}, mismatched~\cite{BarEli1984, Aronson1990}, dynamically coupled ~\cite{Konishi2003}, or nonlinear~\cite{Prasad2010}. Apart from diffusive coupling, the nonlinear coupling is also used in achieving the amplitude death~\cite{Prasad2010, Prasad2003}. In the presence of the nonlinear coupling, the amplitude death occurs in the absence of parameter mismatch and also in the absence of time delay ~\cite{Prasad2010, Prasad2010b}.
 
In this paper, we investigate the migration dilemma in the setup of the CML of replicators' subpopulations and investigate the nonlinear dynamics of the amplitude death induced synchronization. Specifically, we ask the following relevant questions: Firstly, what happens if the interaction between the demes is not simple diffusive coupling but some more general nonlinear coupling? Secondly, does the amplitude death synchronizes the replicator dynamics at the demes that have non-chaotic but periodic dynamics? Thirdly, how is cooperation across the population supported as the degree of the network varies?  Furthermore, costly interactions in networks while studying the coevolution of synchronization and cooperation has gathered a lot of recent interest~\cite{2017_Antonioni_Cardillo_PRL, Rohit_2020_chaos, 2018_Yang_Han-Xin_Zhou}. In such a scenario, the deme---based on the payoff it receives---can decide whether to participate in the process of migration or not. Migration---usually costly for different reasons---is seen in the population of different kind of insects~\cite{Rankin1992}, birds~\cite{2007_Newton}, fishes~\cite{1968_Jones}, and mammals~\cite{Harris2009}. One can find interesting investigations about migration cost in the populations---such as that of insects~\cite{Rankin1992}, white storks~\cite{Flack2016}, and spoonbills~\cite{2015_Tamar_etal, Lok2017}. In the light of these studies, in our setup, another crucial theoretical question can be addressed: How a costly interdemic interaction affects the cooperation levels in the network?  It is of interest for us to understand the effect (if present at all) of such costly migration on averting the migration dilemma via the amplitude death. 

Before we present the results that we have found while investigating the above-mentioned problems, let us first succinctly set up the CML---on which the results are based---in the following section.


\section{The coupled map lattice}
Mathematically, our model comprises of a CML network whose nodal dynamics is governed by a replicator map, which for the two-player--two-strategy games are the most convenient yet nontrivial proving ground for our idea. The one-dimensional replicator map~\cite{1997_BS_JET, 2000_HS_JEE,2000_BV_QJE,2003_C, 2010_M_AEJM, 2011_V_PRL,2018_Pandit_etal_Chaos, 2020_MC_JTB, Archan_2020_chaos} is given by,
\begin{equation}
	x_{n+1}=f(x_n):=x_n+x_n[({\sf A}\mathbf{x_n})_1-(\mathbf{x_n})^T{\sf A}\mathbf{x_n}], \label{eq:RM}
\end{equation}
where subscript `1' denotes the first component of vector ${\sf A}\mathbf{x_n}$, $n$ denotes the time step, and 
\begin{eqnarray*}  
	\label{eq:A1}
	\centering
	\begin{tabular}{cc|c|c|}
		& \multicolumn{1}{c}{} & \multicolumn{2}{c}{{Player $2$}}\\
		& \multicolumn{1}{c}{} & \multicolumn{1}{c}{Cooperate} & \multicolumn{1}{c}{\,\,\,\,Defect\,\,\,\,}\\\cline{3-4} 
		\multirow{2}*{{Player $1$}} & Cooperate & $R$ & $S$ \\\cline{3-4}
		& Defect & $T$ & $P$ \\\cline{3-4} 
	\end{tabular}
\end{eqnarray*}
exhibits the strategies involved and the real-valued payoff matrix ${\sf A}$ for {Player 1} in the two-player--two-strategy symmetric game. ${\bf x}=(x,1-x)$ is the state of the population such that $x$ is the fraction of the cooperators and $1-x$ is the fraction of the defectors. For consistency, it is required that the one-dimensional replicator map be such that $0\le x_n\le1$ for all $n$. This strictly depends on the values of the elements of the payoff matrix. It should be borne in mind that the discrete replicator equation is concerned with the replication-selection across the generations of a vast well-mixed collection of the cooperators and the defectors. 

We judiciously choose the above form of the replicator equation because (a) it phenomenologically models the replication-selection dynamics that is in line with the Darwinian tenet of the natural selection, (b) its fixed points correspond to the Nash equilibria~\cite{1950_N_PNAS} and the evolutionarily stable strategies/states~\cite{1973_SP_Nature} through the folk and related theorems of the evolutionary game theory~\cite{2014_CT_PNAS,2018_Pandit_etal_Chaos}, and (c) most importantly, it is endowed with chaotic attractors~\cite{2011_V_PRL,2018_Pandit_etal_Chaos,Archan_2020_chaos}. Another more popular form of the discrete replicator equation possible~\cite{1978_TJ_MB} that, however, is not conducive to our study as it does not show~\cite{2018_Pandit_etal_Chaos} non-convergent dynamics for the simple two-player--two-strategy symmetric game. While relatively less in vogue in the biological systems, Eq.~(\ref{eq:RM})---as itself or in related forms---also appears in modelling intergenerational cultural transmission~\cite{2000_BV_QJE, 2010_M_AEJM}, boundedly rational players' imitational behaviour in bimatrix cyclic games~\cite{2000_HS_JEE}, and reinforcement learning~\cite{1997_BS_JET}. It is interesting to recall that a good behaviour rule does not require aggregate population behaviour implicit in the natural selection to induce the replicator map; instead, the map is arrived at based on the rational behaviour of the players~\cite{2003_C}.

The basic CML considered in this work is a linear lattice with $N$ lattice sites/nodes and periodic boundary condition such that each lattice site (or in our case deme) is connected to its two nearest neighbours. The individuals interact with each other within the deme via a strategic interaction modelled by the replicator map, and there is migration between connected demes. The dynamics on the CML is given by,
\begin{equation}
	x^i_{n+1}=(1-\epsilon)f(x^i_n)+\frac{\epsilon}{2}[g(x^{i-1}_n)+g(x^{i+1}_n)],
	\label{eq_network1}
\end{equation}
where, the superscript $i$ denotes the $i$th lattice site and $\epsilon$ is the coupling strength measuring the rate of migration to the $i$th node from the $(i-1)$th and $(i+1)$th nodes. We must restrict $\epsilon$ between $0$ to $1$ so that $x^i_n$ does not become either negative or greater than one. In the model under consideration, the migration can be modelled as a linear term or, more generally, as a nonlinear term depending on the function $g(x)$.

We must remark that the coupled continuous replicator dynamics---as opposed to the discrete one used herein---with migration included was studied~\cite{2005_MP_JTB} for a system of two demes in which the individuals play symmetric two-player games such that there exist two ESSs: one payoff dominant (maximal payoff state among the equilibria) and the other risk-dominant (less risk of loss for the players). It was concluded that due to the migration, most of the individuals play the payoff-dominant strategy even if both the demes are initiated in the basin of attraction of the risk-dominant equilibrium. Our system is far richer dynamically since the discrete replicator equation can lead to nonconvergent attractors that are clearly not connected with the game-theoretic concepts like the evolutionary stable strategy (and hence, the Nash equilibrium). Because of the prevalence of non-fixed point dynamics in the dynamical systems, we must address them in spite of rather limited understanding of their connection with the game theory. Needless to say, in the scenario where two or more discrete replicators coupled via migration, evolutionary stable strategies---whether payoff- or risk-dominant---are rendered unachievable in the evolutionary dynamics since these equilibrium strategies, by construction, can only be connected with some convergent fixed point attractors.

\section{Nonlinear coupling}
\label{sec:critical}
\begin{figure}
\includegraphics[scale=0.85]{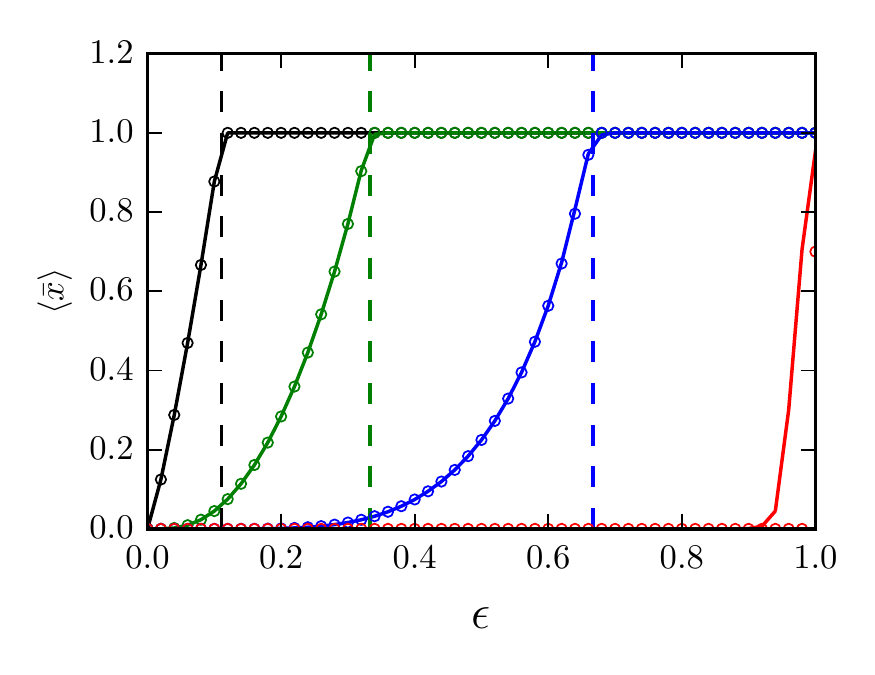}
\caption{\textbf{Emergence of cooperation in the CML with only the PD, nonlinear coupling, and no rewiring.} The average cooperation $\langle \bar{x} \rangle$ at each deme is plotted as a function of coupling strength, $\epsilon$, for the replicator map. The black, green, blue, and red solid lines correspond to the average cooperation as obtained  from the numerical simulations done with $\alpha=0.2$, $0.8$, $0.95,$ and $1.0$ respectively. Circular markers are denoting the homogeneous fixed points for the respective combination of the parameters $\alpha$ and $\epsilon$. The black, green, blue, and red dashed lines are the theoretically predicted critical coupling strength beyond which one gets full cooperation for $\alpha=0.2$, $0.8$, $0.95,$ and $1.0$ respectively. Here we have set $R=1.1,~S=0.0,~T=1.2, ~P=0.1$ in the payoff matrix of PD.}
\label{non_linear_coupling}
\end{figure}

\begin{figure}
	\includegraphics[scale=0.09]{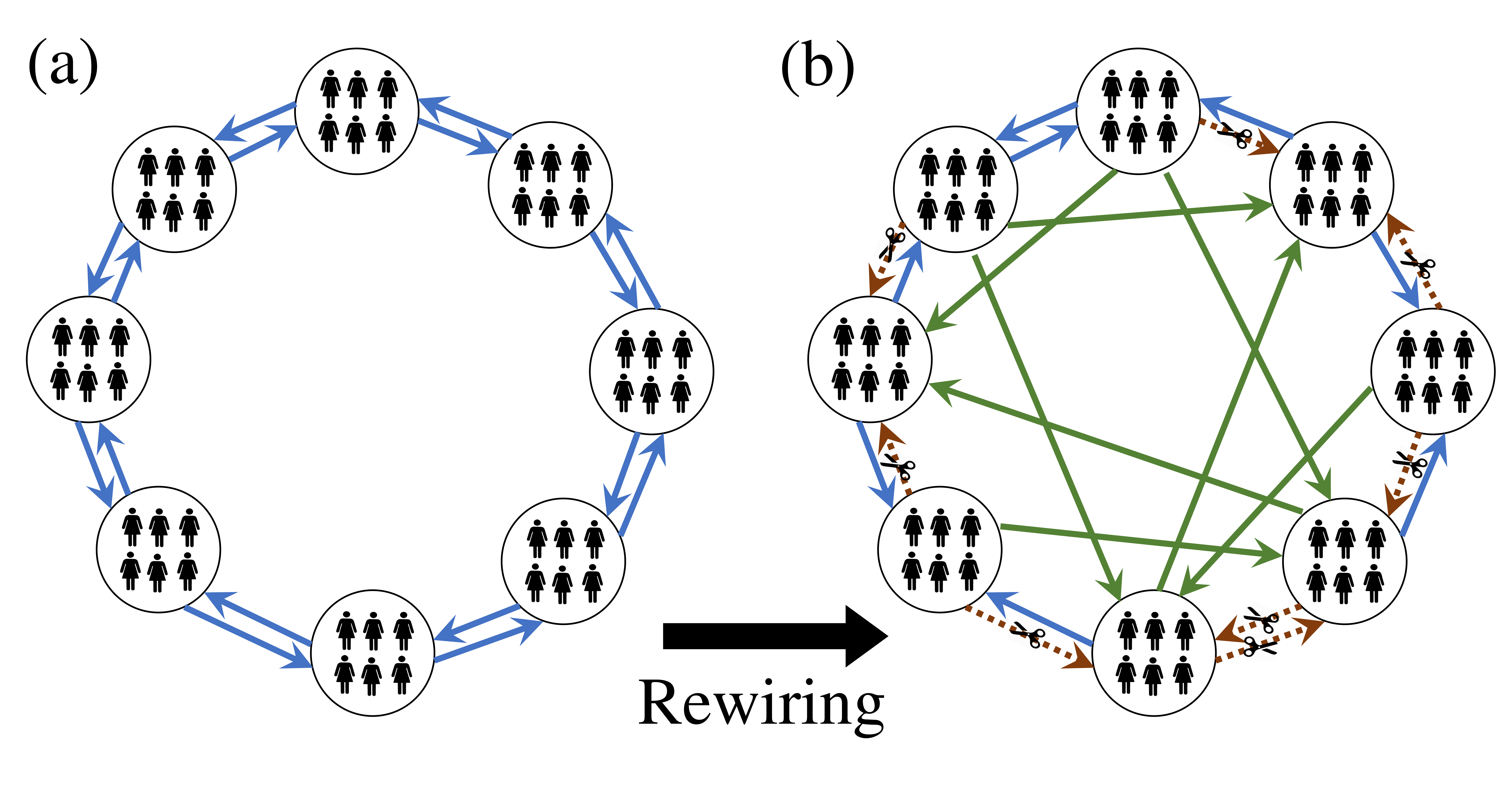}
	\caption{\textbf{Schematic diagram of the coupled map lattice with dynamic random rewiring.} We see in left panel (a) the base CML with eight demes each having six representative individuals for illustrative purpose. Every deme has a game---say, the PD or the LG---played by the individuals in it. The arrowheads point towards the destinations of respective migration. In right panel (b), as the dynamic random rewiring is employed, some of the directed edges (shown by blue arrows) of the base CML are randomly broken (shown by brown arrows with scissors) and new incoming edges (shown by green arrows) are created.}
\label{fig:schematic_diagram}
\end{figure}
%
Now, let us consider a class of nonlinear coupling---a power-law coupling with $\alpha\geq 0$---such that the dynamics of the CML is given by,
\begin{eqnarray}
x^i_{n+1}=f(x^i_n)(1-\epsilon)+\frac{\epsilon}{2}\left[(x^{i+1}_n)^{\alpha}+(x^{i-1}_n)^{\alpha}\right],
\label{eq_network}
\end{eqnarray}
where, $i=1,2,\cdots,N$. This CML has two homogeneous fixed points: $x^*=0$ or $x^*=1$ at all the demes---they, respectively, correspond to all defector or all cooperator states. The choice of power law coupling is mostly to illustrate the possible effects of nonlinear coupling in a mathematical tractable setting. However, a physical motivation for using the power-law coupling may be sought in the seminal work by Zipf~\cite{zipf1949} where such a power law in the growth of city population---due to immigration of people---was predicted. Furthermore, in a recent work~\cite{PrietoCuriel2018} based on US census data, a power law in the migration of people from city to city is observed.

We perform the linear stability analysis~\citep{2002_Sinha_PRE} about a homogeneous fixed point $x^*$ by substituting $x^i_n=x^*+h^i_n$ and expanding the resultant equation up to the first order to get:
\begin{equation}
h_{n+1}^i=\left(1-\epsilon\right)f'\left(x^*\right)h_n^i+\frac{\alpha \epsilon}{2}\left(h_n^{i+1}+h_n^{i-1}\right).\label{pert_alpha}
\end{equation}
Now expressing the small perturbations as a sum of its Fourier components, i.e., $h_n^i=\sum_q \tilde{h}^q_n \exp\left(\sqrt{-1}qi\right)$, where $q$ is the wave number, and substituting in Eq.~(\ref{pert_alpha}), we arrive at the following expression:
\begin{equation}
\frac{\tilde{h}^q_{n+1}}{\tilde{h}^q_n}=f'\left(x^*\right)\left(1-\epsilon \right)+\alpha \epsilon\cos q.
\end{equation}
For perturbation amplitude to decrease with time, i.e., in order for the fixed point to be stable, we need
\begin{equation}
\bigg | \frac{\tilde{h}^q_{n+1}}{\tilde{h}^q_n} \bigg |=\left|f'\left(x^*\right)\left(1-\epsilon \right)+\alpha \epsilon\cos q\right|<1.\label{condition1}
\end{equation}

Therefore, in order to find the condition for establishing full cooperation in the CML when same game is played at all the demes, we necessarily need to find the condition for $x^*=1$ to be stable.  In the case of $0\leq\alpha< 1$,  the critical coupling strength, $\epsilon_{\rm{crit}}$ (the minimum coupling strength required to impart full cooperation), in the CML of replicator maps is given by:
\begin{equation}
\epsilon_{\rm{crit}}=\frac{1}{1+\frac{1-\alpha}{T-R}},
\label{type1}
\end{equation}
where we have explicitly used that fact that for the replicator map, $f'(1)=T-R+1$ . We have verified this result for different values of $\alpha$ using numerical simulations as presented in Fig.~\ref{non_linear_coupling}.

From this expression (Eq.~(\ref{type1})), we can clearly see that if we have the CML where at every deme the prisoner's dilemma (PD) game ($T>R$ and $P>S$) is played, the emergence of cooperation is possible because $\epsilon_{\rm{crit}}$ can be less than unity. However, if $\alpha\geq1$, then the CML with only the PD at each deme can not lead to cooperation because $f'(1)=T-R+1>1$ for the PD and hence we cannot have $x^*=1$ stable (see Eq.~(\ref{condition1})).  
%

\begin{figure}
    \includegraphics[scale=0.57]{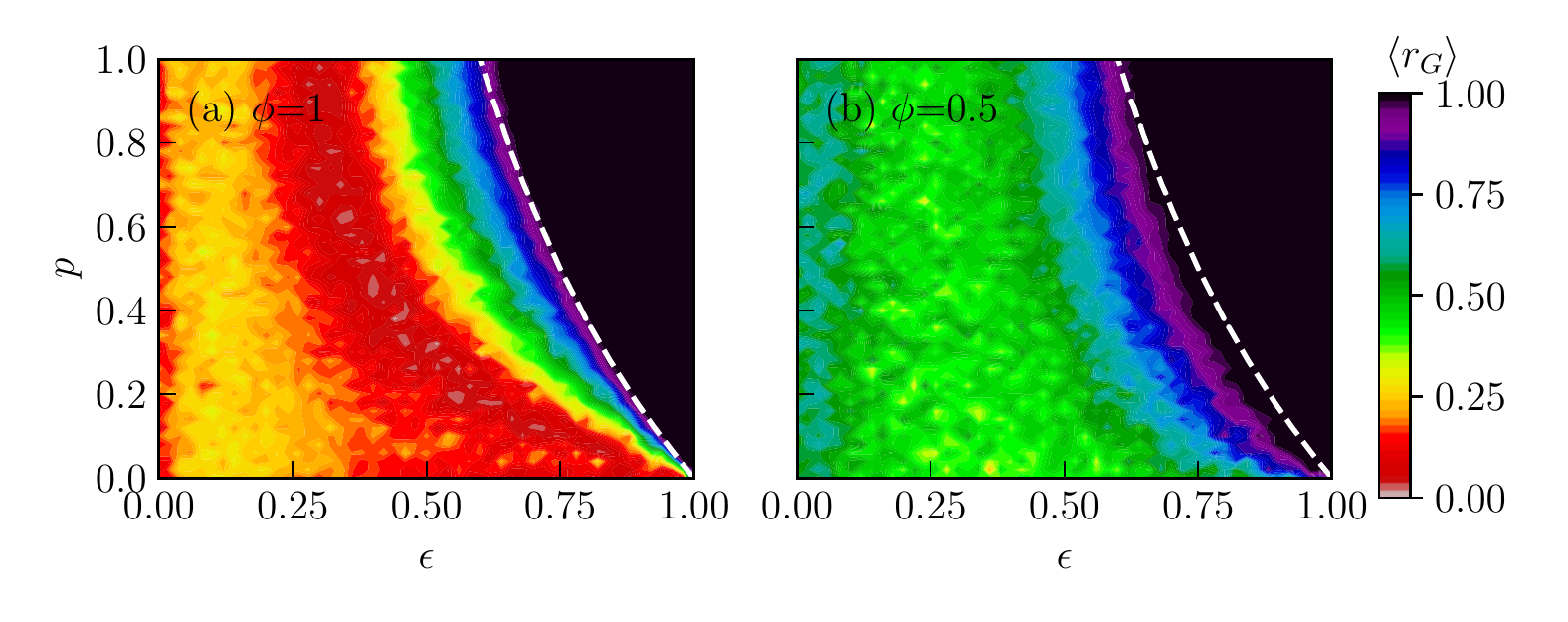}
    \caption{\textbf{Numerical validation of critical coupling strength for the CML with dynamic rewiring.} We show the order parameter, $\langle r_G\rangle$, as a function of coupling strength $\epsilon$ and the rewiring probability $p$. In subplot (a), every deme has the LG, whereas in subplot (b), only half of the demes have the LG (payoff matrix
$\tiny{\begin{pmatrix} 
1 & 7 \\
8 & 0\\
\end{pmatrix}}
$) and the rest have the PD (payoff matrix
$\tiny{\begin{pmatrix} 
1.1 & 0 .0\\
1.2 & 0.1\\
\end{pmatrix}}
$). The white dashed line corresponds to Eq.~(\ref{eq_critical}) specifying how the critical coupling strength varies with the rewiring probability in the CML with exclusively leader games at all demes. The average in- and out-degree of the network is two. Here, $N=100$ demes and the system has been evolved for $2000$ time steps.}
        \label{rewire_crit}
\end{figure}

\begin{figure*}
    \includegraphics[scale=0.7]{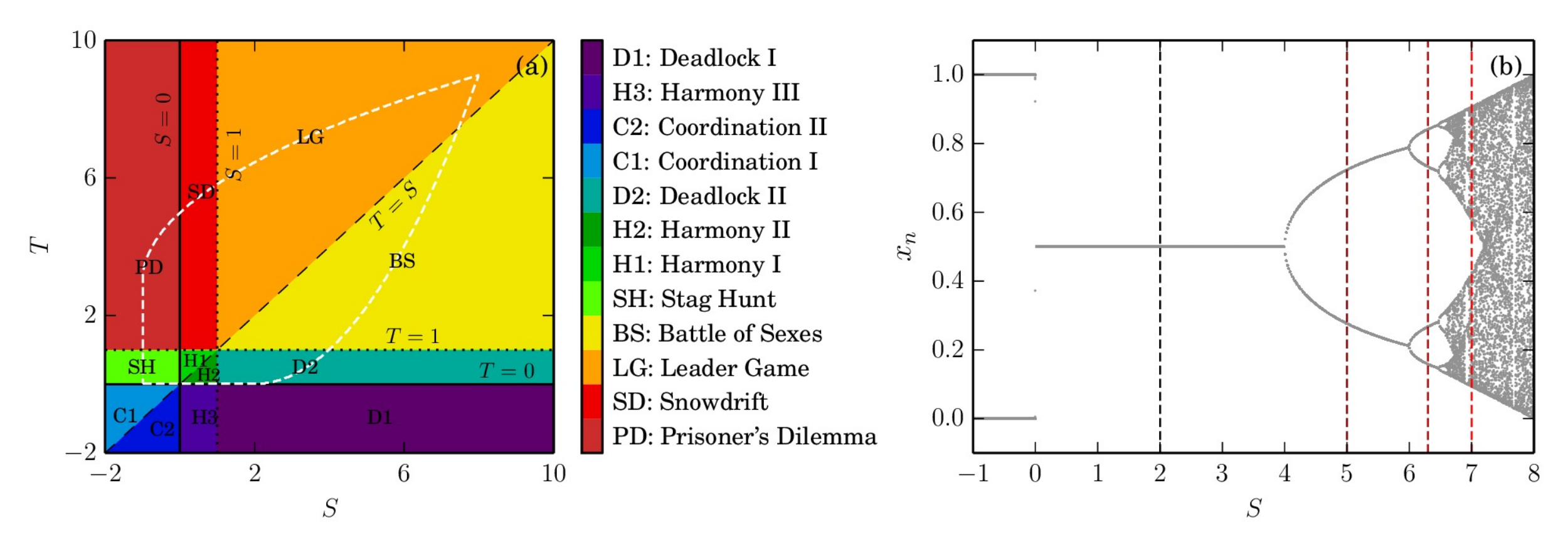}
    \caption{\textbf{Classification of twelve symmetric games.} In subplot (a), we show how the region of the $S$-$T$ parameter space is separated in twelve well-known games whose names have been specified in the figure. The interior of the white dashed leaf-like boundary corresponds to the region where the replicator map has physical solutions. On taking the payoff matrices from the line $T=S+1$ and evolving the replicator map, we find period doubling route to chaos as depicted in subplot (b) where the vertical dashed lines, from left to right, respectively correspond to $\textsf{A}_{\rm P1}$, $\textsf{A}_{\rm P2}$, $\textsf{A}_{\rm P4}$, and $\textsf{A}_{\rm C}$.} 
    \label{fig:leaf}
\end{figure*}
\begin{figure*}
	\includegraphics[scale=0.8]{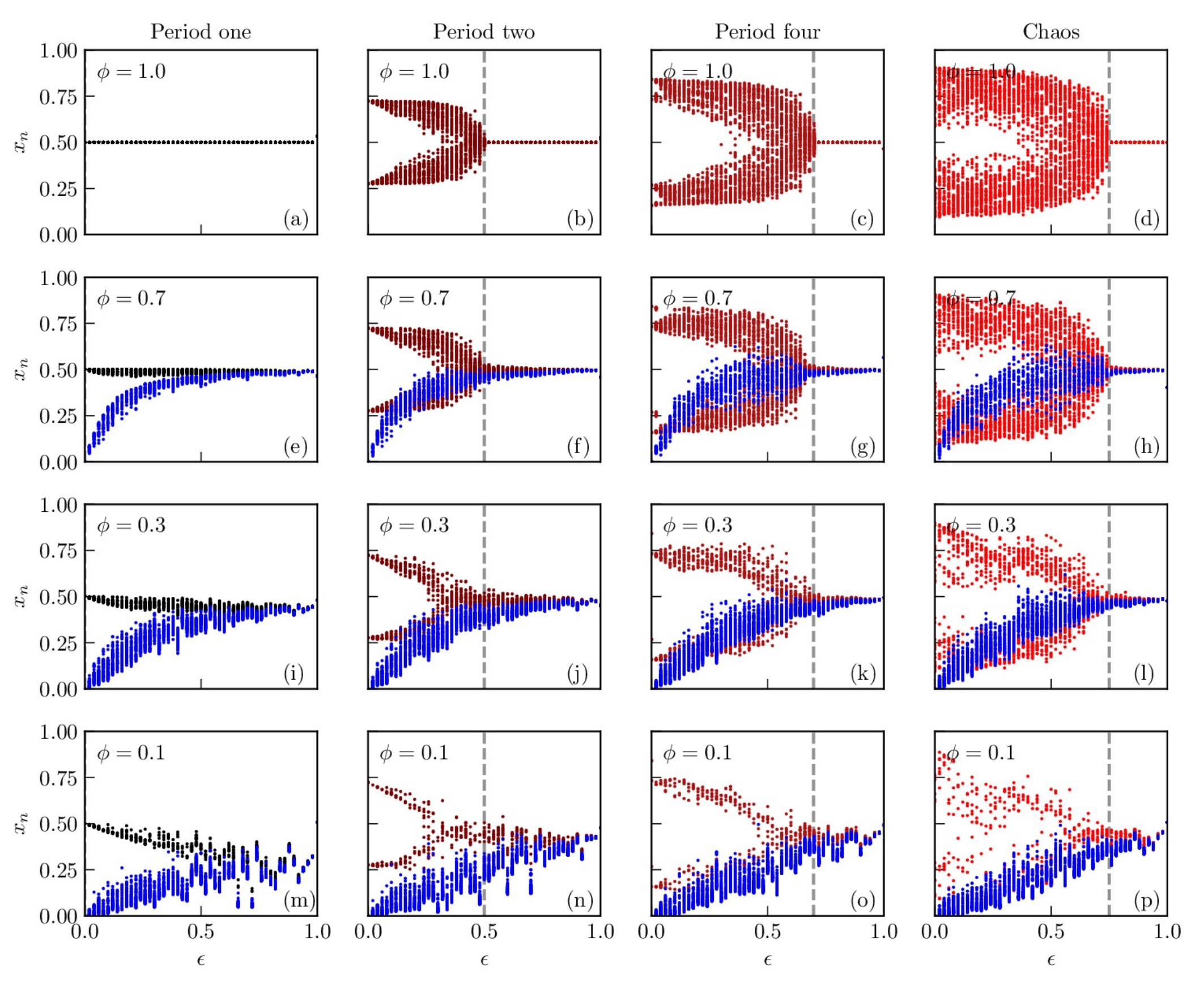}
	\caption{\textbf{Amplitude death: The demes with the LG induce cooperation in the demes with the PD.} When all the demes of the CML are playing the LG with (a) $\textsf{A}_{\rm P1}$, (b) $\textsf{A}_{\rm P2}$, (c) $\textsf{A}_{\rm P4}$, or (d) $\textsf{A}_{\rm C}$ after the respective critical values of  the coupling strength (vertical dashed line) all the demes' trajectories (black, brown, light brown, and red dots respectively for the four aforementioned LGs) synchronize onto the fixed point $x^*=0.5$ of the CML. As we introduce the PD in some of the demes with no (or few) cooperators, then the corresponding trajectories (blue dots) are pulled onto the synchronized state $x^*\approx0.5$ for all demes beyond $\epsilon_{\rm crit}$ as exhibited in rest of the subplots for the LG game fraction, $\phi=0.7$, $0.3$, and $0.1$. Subplots (a), (e), (i), and (m) correspond to $\textsf{A}_{\rm P1}$; subplots (b), (f), (j), and (n) correspond to $\textsf{A}_{\rm P2}$; subplots (c), (g), (k), and (o) correspond to $\textsf{A}_{\rm P4}$; and subplots (d), (h), (l), and (p) correspond to $\textsf{A}_{\rm C}$. Here, $N=100$ demes and the system has been evolved for $2000$ time steps.}
	\label{diff_game_bif}
\end{figure*}
\begin{figure*}
    \includegraphics[scale=0.7]{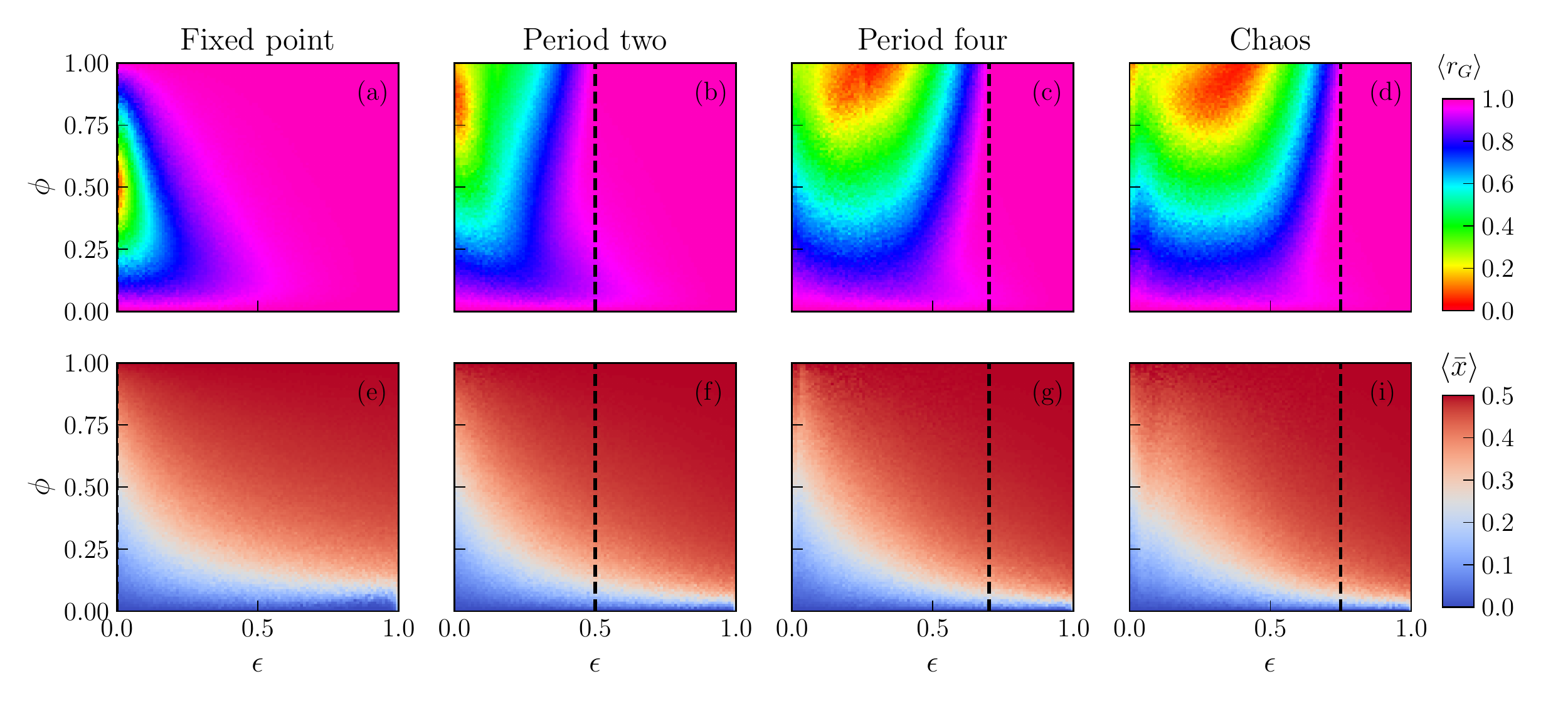}
   \caption{\textbf{Synchronization order parameter and emergence of cooperation.} For the CML with mixed types of demes---some playing the PD and some the LG (with $\textsf{A}_{\rm P1}$, $\textsf{A}_{\rm P2}$, $\textsf{A}_{\rm P4}$, and $\textsf{A}_{\rm C}$ from left to right columns respectively)---we plot order parameter $\langle r_G\rangle$ and average cooperation $\langle \bar{x}_n\rangle$ (averaged over 64 realizations) as functions of coupling strength $\epsilon$ and the leader game fraction $\phi$ for the cases where the LG has period one (or fixed point) attractor ((a) and (e)), period two attractor ((b) and (f)), period four attractor ((c) and (g)), and chaotic attractor ((d) and (i)) when plugged into the replicator map. The vertical black dashed line corresponds to $\epsilon=\epsilon_{\rm crit}$ when the CML has only the LG at all the lattice points. Here, $N=100$ demes and the system has been evolved for $2000$ time steps.}
    \label{diff_game_coop}
\end{figure*}


\section{Random Rewiring}
\label{sec:linear}


Evidently, in the CML consisting of only one kind of demes---each having individuals playing the prisoner's dilemma game and linear coupling ($\alpha=1$) among them, the emergence of cooperation ($x^*\ne0$) requires dynamic rewiring as a possible mechanism~\cite{2002_Sinha_PRE}. In order to implement the random migration in the system under investigation, we modify the couplings in the CML. At every time step, any node can either allow migration from its two nearest neighbours or two other demes picked randomly from an uniform distribution. The probability of remaining coupled to the nearest neighbours is $1-p$, where $p$ is called the dynamic rewiring probability; `dynamic' emphasizes that the rewiring is happening at every time step. 

Mathematically, the mean-field equation for the CML with the random coupling is given by,
\begin{eqnarray}
x^i_{n+1}=(1-\epsilon)f(x^i_n)+\frac{(1-p)\epsilon}{2}(g(x^{i-1}_n)+g(x^{i+1}_n))\nonumber\\+\frac{p\epsilon}{2}(g(x^{\xi}_n)+g(x^{\eta}_n)),
	\label{eq_rewired_network}
\end{eqnarray}
where, $\xi$ and $\eta$ are the indices of the two randomly chosen demes and are not equal to $i-1$, $i$, or $i+1$. Note that while the in-degree of every node remains fixed at two, the out-degree of every node may not remain two. See Fig.~\ref{fig:schematic_diagram}{(b)} for a schematic representation.  We would like to spell out that, in effect, there are two edges---incoming and outgoing---between two nearest neighbours (see, e.g., Fig.~\ref{fig:schematic_diagram}{(a)}). From a deme's perspective, an incoming edge indicates immigration and an outgoing one denotes emigration. The parameter $p$ in Eq.~(\ref{eq_rewired_network}) can also be interpreted as a measure of how strong the long-range migration (or non-nearest neighbour migration) is compared to the short-range migration (nearest neighbour migration) since $\epsilon p/\epsilon(1-p)$---the ratio of the coefficients in the third and the second terms in the right hand side of Eq.~(\ref{eq_rewired_network})---is a monotonically increasing function of $p$ in the range $0$ to $1$.

In the model, there is a possibility that the dynamics at all the demes may be completely synchronized to an interior fixed point, i.e., $x^i=x^j=x^*$ for all $i$ and $j$. As done in the immediately preceding section (Sec.~\ref{sec:critical}), one could do a linear stability analysis to find if this synchronized state is at all stable and hence attainable. It can be shown that such a stable state in fact exists when $\epsilon\ge\epsilon_{\rm crit}$ which can be explicitly given as,
\begin{equation}
	\label{eq_critical}
	\epsilon_{\rm{crit}}=\frac{|f'(x^*)|-1}{|f'(x^*)|-1+p}.
\end{equation}
In passing, we note that with $\alpha=1$ in Eq.~(\ref{type1}) and with rewiring probability $p=0$ in Eq.~(\ref{eq_critical}), both of them reduce to the same value of critical coupling strength, viz, unity, as they should.

We quantify the extent to which the system is synchronized by defining a global order parameter~\cite{2016_Wang}, $r_G:=|\sum_{i=1}^N e^{2\pi \sqrt{-1} x^i}|/N$, that asymptotically reaches unity as the system attains complete synchrony. For large $N$ and uniformly distributed $x^i$ in the interval $[0,1]$, we can easily observe that $r_G=0$. Thus, any partially synchronized state have a non-zero value of $r_G$ that is less than unity. The system is completely synchronized when $r_G=1$ that corresponds to the state of the population where $x^i=x^j$ for all $i,j\in\{1,2,\cdots,N\}$. Fig.~\ref{rewire_crit}(a) presents the verification that the critical coupling strength depends on the rewiring probability $p$ exactly as predicted by Eq.~(\ref{eq_critical}) for the Leader game (LG)~\cite{1967_R_BS,2014_Hummert_etal_MBS}---a game with chaotic solutions~\cite{2018_Pandit_etal_Chaos}. What is even more satisfying is that the prediction is quite good even when the CML has mixed kind of demes---half playing the PD and the rest the LG (see Fig.~~\ref{rewire_crit}(b)). In the figure, and henceforth, $\phi\in[0,1]$ denotes the fraction of the demes at which the LG is played.

\subsection{Twelve ordinal class of games: periodic orbits and chaos}
There are twelve ordinally equivalent classes of two-player--two-strategy symmetric normal form games~\cite{2014_Hummert_etal_MBS, 2018_Pandit_etal_Chaos} that can be represented by the following general payoff matrix:
\begin{equation}
{\sf A} = 
\begin{pmatrix}
1 & S  \\
T & 0
\end{pmatrix}.
\label{mat_payoff}
\end{equation}
One must work with only those values of $S$ and $T$ for which $0\le x^i_n\le 1$ for all values of $i$ and $n$. In the $S$-$T$ plane, the twelve classes of games are clearly demarcated by the straight lines: $S=0$, $T=0$, $S=1$, $T=1$, and $T=S$. The physical region for the replicator map given by Eq.~(\ref{eq:RM}) is shown in Fig.~\ref{fig:leaf}(a). 

Without any loss of generality, for the sake of concreteness, we choose $S=-0.1$ and $T=1.1$ in the PD where obviously a player is better off defecting irrespective of what the opponent chooses from the two strategies available to her. We note that this form of the PD corresponds to the additive model studied by W.~D.~Hamilton~\cite{HAMILTON_1964} and R.~Trivers~\cite{1971_Trivers}. We have already seen that the replicator map corresponding to the PD game displays solutions that asymptotically reach a fixed point attractor. Also, when some of the demes playing the PD, transform into the LG---which has chaotic solutions---the entire CML tries to synchronize to impart cooperation in all the demes. Historically~\cite{1967_R_BS}, the LG's name comes from the fact that if a player shifts its strategy from cooperation to defection, it rewards both the players but herself more so, and hence, is a leader. The LG has a symmetric mixed Nash equilibrium~\cite{1950_N_PNAS} that is an unstable fixed point of the corresponding replicator map.

In order to explicitly see that the PD with altruism rewarded is the LG, we look at the following two specific payoff matrices where subscripts indicate that the matrices are respective that for the PD and the LG:
\begin{equation*}
{\sf A}_{\rm PD} = 
\begin{pmatrix}
1 & -0.1  \\
1.1 & 0
\end{pmatrix}\,{\rm and}\,\,
{\sf A}_{\rm LG} = 
\begin{pmatrix}
1 & -0.1+1.15  \\
1.1 & 0
\end{pmatrix}.
\end{equation*}
It is explicitly evident that on giving a cooperator playing against a defector some extra reward, say, $+1.15$ in the above, can turn the PD game to the LG (as now $T>S>1>0$). The LG can have different entries in the matrix as long as the required defining condition is satisfied. It is interesting to note that as one varies the parameters along the line $T=1+S$, starting from $(S, T)=(-1,0)$, the replicator map undergoes period-doubling route to chaos as shown in Fig.~\ref{fig:leaf}(b). Hence, it is natural to be curious about how other the LGs, whose replicator dynamics lead to other non-chaotic asymptotic solutions, affect the cooperators in the CML with most demes playing the PD. 

To this end, we choose three LGs (marked in Fig.~\ref{fig:leaf}(b)) with following payoff matrices:
\begin{equation*}
{\sf A}_{\rm P1} = 
\begin{pmatrix}
1 & 2  \\
3 & 0
\end{pmatrix},\,\,
{\sf A}_{\rm P2} = 
\begin{pmatrix}
1 & 5  \\
6 & 0
\end{pmatrix}, \,{\rm and}\,\,
{\sf A}_{\rm P4} = 
\begin{pmatrix}
1 & 6.3  \\
7.3 & 0
\end{pmatrix},
\end{equation*}
that respectively correspond to replicator dynamics with asymptotically stable period one (fixed point) orbit, period-two orbit, and period-four orbit. We set parameters $S=7$ and $T=8$ to choose another LG (also marked in Fig.~\ref{fig:leaf}(b)) endowed with a chaotic attractor. We denote its payoff matrix by $\textsf{A}_{\rm C}$

\subsection{Mixed games in demes}
We set the rewiring probability to $p=0.5$. If the CML only has an LG with fixed payoff matrix at all the demes then synchronization ($r_G\approx1$) onto $\langle \bar{x}\rangle\approx0.5$ occurs at threshold critical coupling parameters that respectively are $0$, $0.5$, $0.7$, and $0.75$ for $\textsf{A}_{\rm P1}$, $\textsf{A}_{\rm P2}$, $\textsf{A}_{\rm P4}$, and $\textsf{A}_{\rm C}$. It should be noted that this effectively is amplitude death because all the periodic or chaotic oscillations die when the synchronized state is reached.

 Just like the case of mixed demes with the PD and the chaotic LG (with ${\sf A}_{\rm C}$)~\cite{2021_Sadhukhan_PRR}, if the chaotic LG is replaced by any of the other aforementioned LGs, one can see that, starting from infinitesimal cooperator fraction, enough cooperation can be established in the demes playing the PD if the random migration is strong enough. Fig.~\ref{diff_game_bif} exhibits this fact in detail and transparently. The symmetry in the subplots of the figure about the line $x=0.5$ is easily explained. Along the line $T=1+S$ in the parameter space if one recasts Eq.~(\ref{eq_rewired_network}) using $x_n^i\rightarrow y_n^{i}=x_n^i-0.5$, the dynamics in terms of $y$-variable---whose range is $[-0.5,0.5]$ for all $i$---appears to be symmetric about zero: $y_{n+1}=y_{n}+{S}y_n\left(4{y_n}^2-1\right)/2$.
 
The emergence and sustenance of cooperation is further depicted in Fig.~\ref{diff_game_coop}. Note that remarkably high degree of synchronization---viz., $\langle r_G\rangle\approx1$, angular brackets denoting average over many realizations of random migration---is seen for the very low values of $\phi$ for all coupling strengths and for coupling strength more than a critical value for almost all $\phi$. The reason for the former is that the demes with the PD dominate and defectors cannot be replaced, and the reason for the latter is that the synchronization of the chaotic LG pulls the cooperator fraction of all the demes with the PD up. The high values of sustained cooperation---viz, $\langle \bar{x}\rangle\approx0.5$, overbar denoting the average over demes---owe to the high values of coupling strength (mostly around and beyond $\epsilon_{\rm crit}$) for almost all $\phi$ (except for very low values). Note that $\langle \bar{x}\rangle\approx0.5$ at intermediate and lower values of $\epsilon$ is accompanied with a low degree of synchronization, meaning that at any given instant cooperator-fraction at a given deme can either be lower or higher than $\langle \bar{x}\rangle$ and this averages out to give the high values of cooperation. 
\section{varying degree}
\label{v_deg}
In the CML that we have considered so far, the in-degree of every node is always two even when the dynamic rewiring is in action. However, one could think of a situation where the in-degree is more than two: A straightforward generalized scenario would be when the in-degree of every node is same $k\in\{2,4,6\cdots\}$ such that each node can be connected with another node at most once and every stage of rewiring is done with probability $p$. We conveniently choose $k$ to be an even number so that in absence of any rewiring each node has $k$ incoming edges---one each from its nearest $k/2$ neighbours on each side. The mean field equation for node $i$ is,
\begin{equation}
    x_{n+1}^i=f(x_n^i)(1-\epsilon)+\frac{\epsilon (1-p)}{k}\sum_{\substack{l=-k/2\\ l\neq 0}}^{k/2} x_n^{i+l}+\frac{\epsilon p}{k}\sum_{l=1}^{k} x_n^{\xi_l}.
    \label{degree_linear}
\end{equation}
$\xi_l$ are the randomly chosen neighbours other than the node $i$ and its nearest $k/2$ neighbours on the either sides. For notational convenience, we denote the $k/2$ neighbours on the clockwise direction by $i+l$ where $l=1,2,...,k/2$ and on the anticlockwise direction by $i-l$.

Assuming that an interior fixed point exists such that at each deme $x=x^*$, we find its stability by putting $x^i_{n+1}=x^*+h^i_{n+1}$ in Eq.~(\ref{degree_linear}), and subsequently keeping only up to the linear order terms for the perturbations. We arrive at
\begin{equation}
\label{perturb_degree}
    h^i_{n+1}=(1-\epsilon)f'(x^*)h^i_{n}+\frac{\epsilon(1-p)}{k} \sum_{\substack{l=-k/2\\ l\neq 0}}^{k/2} h_n^{i+l}.
\end{equation}
Note that we have neglected the randomly changing neighbours' contribution to this equation because that should average out to zero. Again, we make the ansatz: $h^i_{n}=\sum_q{\tilde{h}^q_{n}e^{\sqrt{-1}qi}}$, and Eq.~(\ref{perturb_degree}) yields,
\begin{equation}
\label{condition_degree}
    \frac{\tilde{h}^q_{n+1}}{\tilde{h}^q_{n}}=(1-\epsilon)f'(x^*)+\frac{2\epsilon(1-p)}{k}\sum_{l=1}^{k/2} {\rm{cos}}~ lq.
\end{equation} 
A stable fixed point, or in other words, a stationary synchronous state (akin to ampplitude death in coupled oscillators), exists when the perturbations die out, i.e., modulus of the right-hand side in Eq.~(\ref{condition_degree}) is less than unity. 

\begin{figure}
	\includegraphics[scale=0.63]{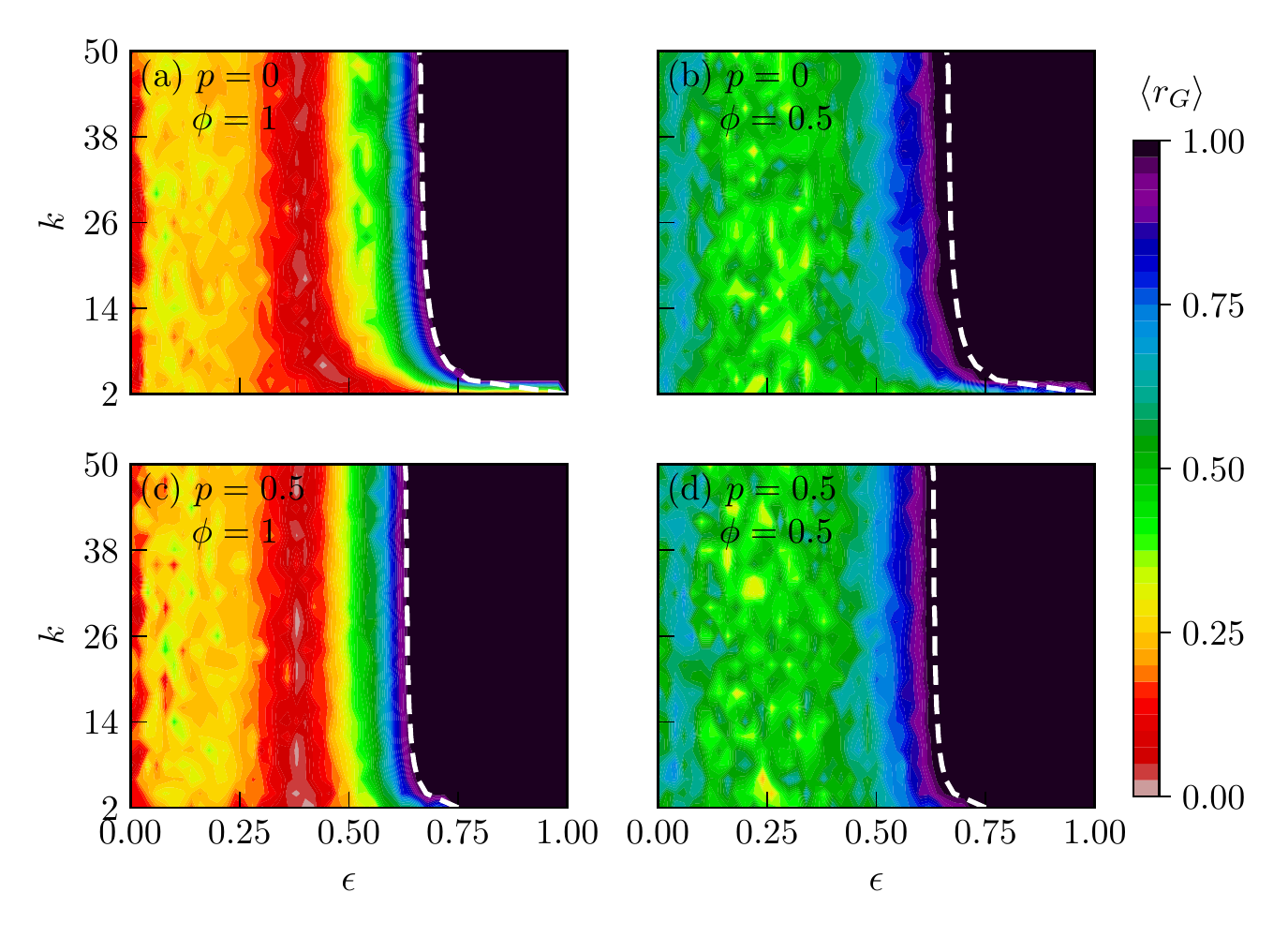}
	\caption{\textbf{Amplitude death induced cooperation saturates with degree.}  We plot the order parameter $\langle r_G\rangle$ as a function of the coupling strength $\epsilon$ and the in-degree of the nodes of the network $k$. In subplots (a) and (c), every deme in the network has the LG (with ${\sf A}_{\rm C}$), while in subplots (b) and (d) only half of the nodes have the LG (the rest have the PD). Viewed from the perspective of the rewiring probability, subplots (a) and (b), and subplots (c) and (d) respectively correspond to $p=0$ and $p=0.5$. The white dashed lines exhibit the analytically estimated critical coupling strength (beyond which amplitude-death induced synchronization is expected) as a function of the in-degree for the case where the network has the LG exclusively. Here, $N=100$ demes and the system has been evolved for $2000$ time steps.} 
	\label{fig_critical_degree}
\end{figure}

The critical coupling strength beyond which it so happens can easily be found numerically, as shown in Fig.~\ref{fig_critical_degree}. We remark that although the above calculations are for the case of single kind of demes in the CML, it gives close enough estimates even when we have mixed kinds of demes---some playing the PD and some playing the LG as evident from Figs.~\ref{fig_critical_degree}(b) and~\ref{fig_critical_degree}(d). The most important outcome of increasing degree is that it can overcome the necessity of random wiring for effecting cooperation through amplitude death (contrast the upper row with the lower row in Figs.~\ref{fig_critical_degree}). However, the effect of degree saturates quite fast; we note that the figures are almost identical beyond $k\approx10$. Since we have already seen in the preceding section that all types of LGs have effectively similar qualitative outcomes, here we have illustrated our results only for the chaotic LG; the degree-dependece of the other LGs are qualitatively similar as expected.  

\begin{figure*}
\includegraphics[scale=0.85]{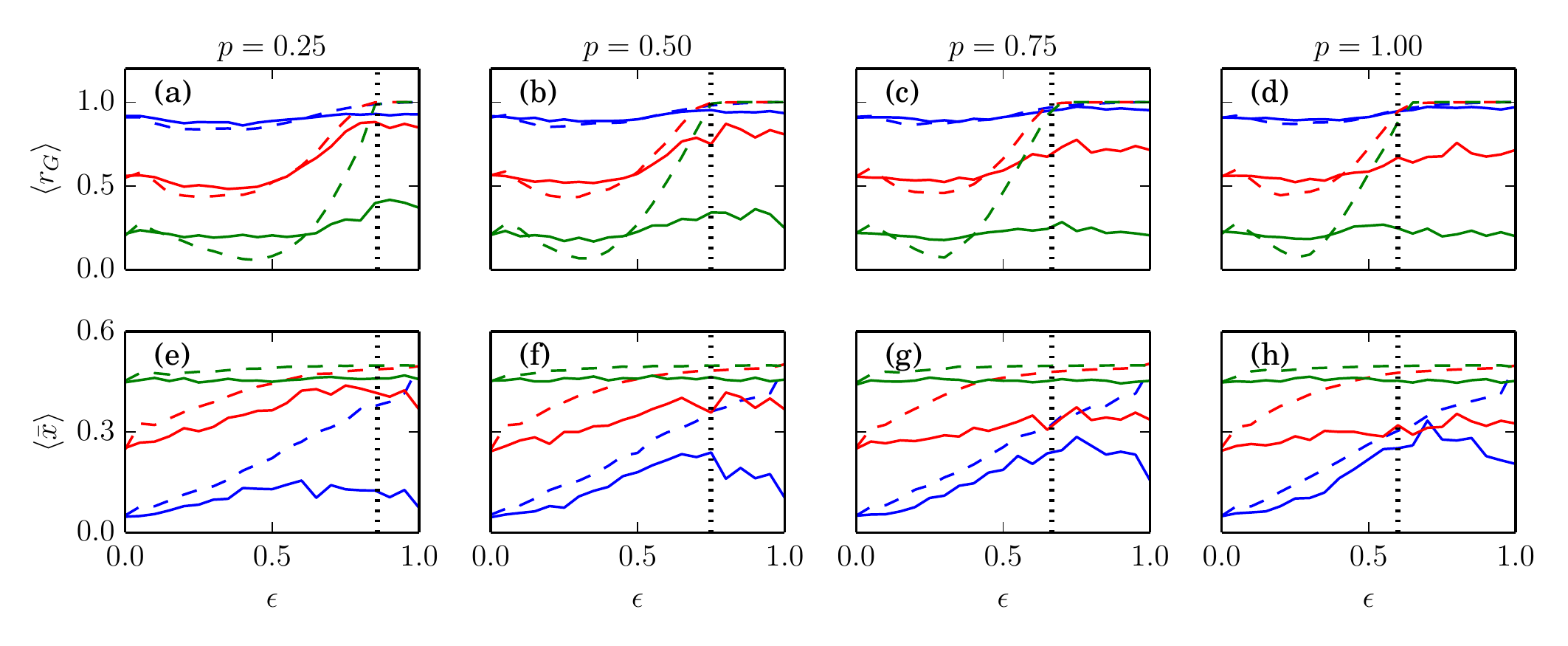}
\caption{{\bf Effect of $\phi$ on synchronization and intrademic cooperation.} The order parameter $\langle r_G\rangle$ ((a)-(d)) and the average intra-cooperation $\langle \bar{x}\rangle$ ((e)-(h)) is plotted for the entire range of the coupling strength $\epsilon$ for a fixed initial cooperating deme fraction  ($C_0=0.1$).  The four columns present the simulation results for the rewiring probability, $p=0.25,~p=0.5,~0.75,$ and $1.0$ respectively. Blue, red, and green colours correspond to $\phi=0.1,~0.5, ~{\rm and}~0.9$ respectively. The dashed lines are used for the case of $C_0=1.0$ where strategy update and hence, costly interdemic interactions is effectively absent. Here, we use $N=100$ demes and evolve the system for $2000$ time steps. All the results are averaged over $512$ realizations. The critical coupling strength for case $\phi=1$ is shown by black dotted lines.}
\label{fig_effect_of_phi}
\end{figure*} 
\begin{figure*}
\includegraphics[scale=0.75]{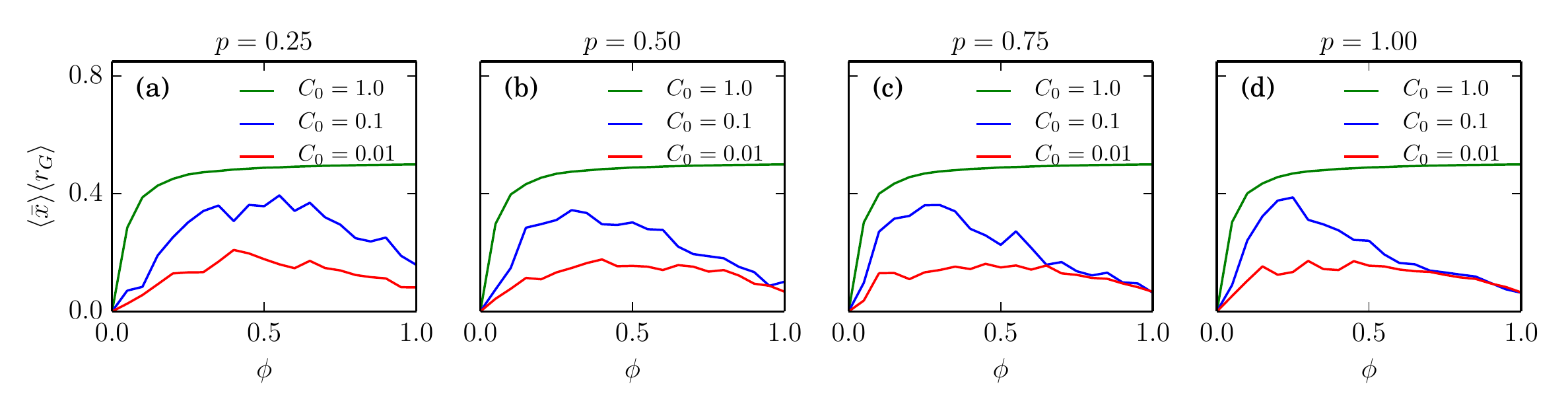}
\caption{{\bf Optimal region for the coevolution of intrademic cooperation and synchronization.} The degree of coevolution of intrademic cooperation and its synchronization $\langle \bar{x}\rangle \langle r_G\rangle$ is plotted with the fraction, $\phi$, of demes where cooperation is rewarded. We fix the coupling strength, $\epsilon=0.9$ and study for different rewiring probability, $p=0.25,~0.5,~0.75,~{\rm and}~1.0$ respectively, in four different columns. In each subplot, we have three different curves corresponding to three different initial inter-cooperation levels, viz., $C_{0}=1.0, ~0.1,~{\rm and}~0.01$.  All the results are averaged over 512 realizations.}
\label{fig_effect_of_p}
\end{figure*}
\section{Strategic interdemic interaction}
\label{strategic_interaction}
As mentioned in the introduction of this paper, migration is known to be costly~\cite{Rankin1992,Flack2016,2015_Tamar_etal, Lok2017}. The subpopulations may decide whether to cooperate by participating in the migration or to avoid migration because the migration is costly. Here we study the effect of costly migration on intrademic cooperation and synchronization in our model, where cooperation is established through random migration. As done in the case of the evolutionary Kuramoto dilemma~\cite{2017_Antonioni_Cardillo_PRL}, we introduce a strategic interdemic interaction that models whether the demes participate in migration based on their acquired payoffs. Note that this effectively introduces an idea of cooperation at the level of demes on top of the idea of cooperation between the replicators of any particular deme.

Mathematically, we modify Eq.~(\ref{eq_rewired_network}) to incorporate the costly interdemic interaction:
\begin{eqnarray}
x^i_{n+1}=f(x_n^i)(1-s^i_n\epsilon)+\frac{s^i_n\epsilon(1-p)}{2}(x^{i+1}_n+x^{i-1}_n)\qquad\nonumber\\+\frac{s^i_n\epsilon p}{2}(x^{\xi}_n+x^{\eta}_n).~~~~~~
\label{eq_interdemic}
\end{eqnarray}
Here $s^i_n$ is the action of the $i$th deme at the $n$th time step; it can take binary values---zero and one. From Eq.~(\ref{eq_interdemic}) we can see if $s^i_n$ is zero then the $i$th deme does not participate in migration. The demes with strategy $s^i=0$ are not cooperators; they do not participate in migration to set up a fixed synchronized level of intra-cooperation in the entire network through amplitude death. On the other hand, demes with $s^i=1$ are coupled to other demes through migration; they may be said to be cooperating in trying to establish cooperation throughout the entire network of the subpopulations. Therefore, while the intrademic cooperation level is quantified by $\langle {\bar x}\rangle$, the interdemic cooperation level at $n$th time step can conveniently be defined as,
\begin{eqnarray}
C_n\equiv\frac{1}{N}\sum_{i=1}^N s^i_n.
\label{eq_interdemic_cooperation}
\end{eqnarray}
We are interested in this section to see how $\langle {\bar x}\rangle$ and $r_G$ depend on $C_0$---the initial interdemic cooperation fraction.

Of course, in order for $s_n^i$ to be a non-trivial game theoretic action, one has to associate some payoffs for the corresponding player. In the present context, it is straightforward because on choosing to cooperate (or equivalently, participate in migration), the deme has to pay a cost that we take as the rate of the deviation~\cite{2017_Antonioni_Cardillo_PRL,Rohit_2020_chaos} from its state:
\begin{eqnarray}
c_n^i\equiv| [x^i_n-f(x_{n-1}^i)]-[x^i_{n-1}-f(x_{n-2}^i)]|.
\label{eq_cost}
\end{eqnarray}
Obviously, the demes with $s^i=0$ incur no cost. We quantify the benefit of each deme as a measure of how much in synchrony it is with its neighbours, and hence, it is aptly measured by the local order parameter~\cite{2017_Antonioni_Cardillo_PRL, Rohit_2020_chaos} given by,
\begin{eqnarray}
b^i\equiv\frac{\sum_j r_{ij}a_{ij}}{\sum_j a_{ij}}.
\end{eqnarray}
Here, $r_{ij}\equiv\frac{1}{2}|(e^{2\pi\iota x_i}+e^{2\pi\iota x_j})|$ (with $\iota=\sqrt{-1}$) is the pairwise order parameter. The quantity $a_{ij}$ is adjacency matrix that takes value $1$ if $i$th and $j$th demes interact through migration; otherwise, it is zero. Therefore the benefit is nothing but the average pairwise order parameter over the neighbours of the deme. In conclusion, the total payoff acquired by a deme should be given by, $U^i\equiv b^i-\alpha c^i$, where $\alpha$ is the relative cost modulating the effect of the cost.

Like in all evolutionary games, the actions $s^i_n$ must also evolve in time in accordance with an update rule. In some of the simplest possible update rules, any focal player (deme, in our context) would compare (in some way) its payoff with that of its neighbours to decide whether to change its action in the subsequent time step so as to reap more payoff. This strategy update can be done stochastically using the myopic rationality factor $\beta$, such that the probability of changing the strategy $s^i$ of $i$th deme to the strategy of its neighbour (say, $s^j$ of $j$th deme) is given by $P_{s^i\rightarrow s^j}=(1+e^{\beta(U^i-U^j)})^{-1}$~\cite{1993_B_GEB,PhysRevE.58.69,2009_RCS_PLR}. In the analogous deterministic rule ($\beta\rightarrow \infty$, effectively), the focal player simply imitates the most successful strategy of its neighbours; since the deterministic rule alters no result qualitatively, we exclusively use the deterministic rule for the strategy update in this paper.

We simulate Eq.~(\ref{eq_interdemic}) with $N=100$ demes and employ the aforementioned deterministic update rule. In general we work with inhomogeneous demes meaning that $\phi$ is a variable in our simulations. For PD, we set $T=1.1$ and $S=-0.1$, whereas we set $T=8,~S=7$ for the demes with the (chaotic) LG. We start our simulations with random initial intrademic cooperation. We keep $\alpha=0.01$ in our simulations. The initial actions $s^i_0$ for the interdemic interaction is assigned randomly such that the initial interdemic cooperation fraction $C_0$ is fixed to some pre-decided small value. We have seen in preceding sections that the random migration helps to synchronize the intrademic cooperation beyond a critical coupling strength $\epsilon_{\rm crit}$ in the absence of interdemic interaction. It is clear from the strategy update rule that the case $C_0=1$ effectively corresponds to the absence of any effect of costly interdemic interaction at any point of time because the only strategy available is to cooperate. 
 
As presented in Fig.~\ref{fig_effect_of_phi}, we numerically find contrasting effects of the costly interdemic interaction on the intrademic cooperation and the synchronization as $\phi$ changes. We fix initial interdemic cooperation fraction as $C_0=0.1$ for concreteness; and we contrast the results found in this case with the case when the costly interdemic interaction (or in other words, the strategy update rule) is absent (realized in the numerical simulation by putting $C_0=1$). 

We note that some general gross features are independent of $p$: The degree of synchronization (see Figs.~\ref{fig_effect_of_phi}(a)-\ref{fig_effect_of_phi}(d)) is lower for higher fraction ($\phi$) of demes with the chaotic LG, whereas the intrademic cooperation level (see Figs.~\ref{fig_effect_of_phi}(e)-\ref{fig_effect_of_phi}(h)) is higher for higher $\phi$. This can be attributed to the fact that when interdemic cooperation level is not unity, it effectively means that some demes are isolated, i.e., they do not participate in the interdemic interaction (or migration); therefore, higher values of $\phi$ means possibility of higher fraction of isolated demes with the LG that ensures some intrademic cooperation but---being chaotic and uncoupled---asynchronous dynamics. Moreover, the intrademic cooperation level is always lower (at any $p$, $\phi$, or $\epsilon$) when strategic interdemic interactions are in action than when they are absent (see Figs.~\ref{fig_effect_of_phi}(e)-\ref{fig_effect_of_phi}(h)). This is again because the isolated demes with the LG cannot interact with the demes with the PD to effect higher cooperator-fraction in them. For similar reasons, we also observe that above $\epsilon_{\rm crit}(p)$ (the critical coupling strength at $\phi=1$ and $C_0=1$ as a function of $p$)---at which we expect synchronized dynamics (hence, $\langle r_G\rangle\rightarrow1$) in the absence of any interdemic strategy update rule---the degree of synchronization is lower when strategic interdemic interactions are in action than when they are absent (see Figs.~\ref{fig_effect_of_phi}(a)-\ref{fig_effect_of_phi}(d)).

Obviously, high values of $\phi$ are good for intrademic cooperation but not as good for synchronization, whereas low values of $\phi$ are good for synchronization but not good for intrademic cooperation. Thus, as hinted by Fig.~\ref{fig_effect_of_phi}, we expect an optimal regime in the parameter space of $\phi$, where both the intrademic cooperation and the synchronization are supported. We quantify the degree of this co-evolution by the product $\langle r_G\rangle\langle \bar{x}\rangle$ which should be low if either $\langle r_G\rangle$ or $\langle \bar{x}\rangle$ is very low. We plot $\langle r_G\rangle\langle \bar{x}\rangle$ versus $\phi$ in Fig.~\ref{fig_effect_of_p} for various initial interdemic cooperation level, $C_0$, including for $C_0=1$ which corresponds to the permanent absence of any effect of costly interdemic interaction. We do note that---irrespective of the rewiring probability $p$---the higher is the initial costly interdemic cooperation, the higher is the degree of co-evolution. 

To understand the nature of the curves in Fig.~\ref{fig_effect_of_p}, we first consider the trivial case of $C_0=0$ which corresponds to the absence of any interdemic interaction at all, costly or otherwise; in fact, even at any later time, the interdemic cooperation level remains zero because the only available strategy is defection. Thus, for the case of mixed (isolated) demes (with the PD and the LG) in the CML, $\langle \bar{x}\rangle$ and $\langle r_G\rangle$ vanish for $\phi=0$ and $\phi=1$ respectively. With increase in $\phi$, one expects $\langle \bar{x}\rangle$ to increase but $\langle r_G\rangle$ to decrease as the number of demes with the LG increases; thus, clearly, one expects a maximum for $\langle r_G\rangle\langle \bar{x}\rangle$ because $\langle r_G\rangle\langle \bar{x}\rangle$ is zero at the two end points, $\phi=0$ and $\phi=1$. With increase in the value of $C_0$, some interdemic connections are established. At $\phi=0$, one still has zero intrademic cooperation as only the PD is present. However, at $\phi=1$, because of the interdemic couplings introduced, some synchronization ($\langle r_G\rangle\ne0$) is established beyond $\epsilon_{\rm crit}$ (this is the reason we choose high $\epsilon$ value in Fig.~\ref{fig_effect_of_p} for illustrating our point). Consequently, $\langle r_G\rangle\langle \bar{x}\rangle\ne0$ at $\phi=1$ although $\langle r_G\rangle\langle \bar{x}\rangle=0$ at $\phi=0$ since the demes with the PD do not allow for any intrademic cooperation ($\langle \bar{x}\rangle=0$) even in the presence of interdemic cooperation. The less is the set of isolated demes (higher $C_0$), the more is the degree of synchronization; and hence, the more is the degree of co-evolution of cooperation and synchronization. One sees this fact validated in Fig.~\ref{fig_effect_of_p} when one compares the two curves corresponding to $C_0=0.01$ and $C_0=0.1$. Of course, $C_0=1$ leaves no demes isolated and hence establishes full synchrony for $\phi=1$, leading to high value of intrademic cooperation as well.

In summary, the take-home message from the studies presented in this section would be the following: The co-emergence and sustenance of the intrademic cooperation and synchronization---effected by incentivizing the cooperators in a few demes with the PD---is hampered in the presence of costly interdemic interactions.

\section{Summary}

The interplay between emergence of cooperation and synchronization is an intriguing phenomenon and a recent fertile direction of research in the evolutionary game theory. In this paper, we have put forward our insights on this interplay by using the coupled map lattice of replicator maps where nonconvergent outcomes, like the periodic orbits and the chaotic orbits, are present even in the evolutionary games with only two strategies. In the model, the emergence of cooperation occurs as the replicator maps synchronize onto a fixed point---a phenomenon very similar to the amplitude death in coupled nonlinear oscillators. The migration dilemma that arises in the system under consideration presents a well known situation---just like in other social dilemmas such as the tragedy of commons~\cite{Hardin1243}, the prisoner's dilemma~\cite{1965_RC}, and agglomeration dilemma~\cite{2011_Roca_Helbing_PNAS}---where an individual's interest is at odds with the entire population's interest as a whole. We have discussed many factors that can lead to the aversion of the migration dilemma in the model. 

The general form of the model, taking into consideration all the factors investigated in this paper, may be summarized by the following equation:
\begin{eqnarray}
\label{general}
x^i_{n+1}=f(x_n^i)(1-s^i_n\epsilon)+\frac{s^i_n\epsilon(1-p)}{k}\sum_{\substack{l=-k/2\\ l\neq 0}}^{k/2}(x^{i+l}_n)^{\alpha}\qquad\nonumber\\+\frac{s^i_n\epsilon p}{k}\sum_{l=1}^{k} (x_n^{\xi_l})^{\alpha};\,\, (i=1,2,\cdots,N).\qquad
\end{eqnarray}
There are five main parameters in our model: (i) the coupling strength $\epsilon$, (ii) the rewiring probability $p$, (iii) the strategy $s^i_n$ of the $i^{th}$ node at the $n^{th}$ time step, (iv) the degree of the network $k$, and (v) the exponent $\alpha$ that makes the migration nonlinear. The model used in Sec.~\ref{sec:critical} for nonlinear coupling is obtained by fixing $p=0$, $k=2$, and $s^i_n=1\textrm{ }\forall i,n$ in Eq.~(\ref{general}). The model used in Sec.~\ref{sec:linear} is obtained by putting $\alpha=1$, $k=2$, and $s^i_n=1\textrm{ }\forall i,n$ in Eq.~(\ref{general}). Similarly, in Sec.~\ref{v_deg}, the degree $k$ of the network is greater than 2; and $\alpha=1$ and $s^i_n=1\textrm{ }\forall i,n$. Finally, we have introduced costly migration between the demes in Sec.~\ref{strategic_interaction} where we have calculated $s^i_n$ for each node $i$ at each time step $n$ with $k=2$ and $\alpha=1$.

We have shown that the migration modelled with a nonlinear power-law function can impart full cooperation even if the whole population is playing only the PD. Moreover, when the migration is modelled with a linear function, a little initiative in rewarding altruism in a few of the demes---thus, transforming the PD to the LG---goes a long way in raising the cooperator-fraction of the population. The network's degree also plays an essential role as we do not require random migration to establish cooperation across the network when the network's degree is greater than two. If the degree is high enough, then the critical coupling strength for synchronization becomes effectively independent of the network's degree. We have also presented in detail how the interdemic interaction due to the costly migration affects the order of synchrony quite nontrivially. The extent of the simultaneous emergence of cooperation and synchrony depends on the fraction of demes where altruism is rewarded. Interestingly, the magnitude of this simultaneous emergence of cooperation and synchrony is maximum at a finite value (not unity) of the fraction $\phi$. It is also important to note that the interdemic cooperation is required for cooperation to be sustained in the individual subpopulations. Such a multilevel interaction can be thought of as a potential model for studying group selection~\cite{1982_K_E,1982_CA_PNAS,1990_R_AN,2006_TN_PNAS}. 

\section{Discussion}

The random migration, which has been crucial in arriving at the results in this paper,  may be seen in a different light: For the individuals in a particular deme when migrating to another deme, there is a chance to get involved in a different game than what they have been playing before migration. Especially, for the payoff matrices chosen in this paper, a player previously playing the PD when migrates to a deme playing the LG, it gets a chance to reap more payoff. Thus, effectively, the CML with random migration is somewhat akin to a stochastic game~\cite{2018_Hilbe_etal_2018,2019_SMWN_PNAS}; however, it is more intriguing as the fraction of players jumping into a neighbouring deme with different payoff matrix is determined by a nonconvergent deterministic dynamics. In passing, we mention that the fact that the random migration helps in establishing a higher cooperation state is interestingly at odds with what happens in the case of demes with a finite population where an increase in the migration rate, for the same benefit-to-cost ratio, decreases the probability of fixation by the cooperators~\cite{2006_TN_PNAS}. 

We would also like to contrast our migration-based model with other diffusion-based models in the literature. In a model~\cite{2009_WNH_PNAS} with asymmetric diffusion rate for the cooperators and the defectors in a population with the ecological public goods games played in subpopulations leads to pattern formations in the densities of the cooperators and the defectors. When a third strategy, loner---agent who does not participate in public goods game---is incorporated~\cite{Valverde2017} in the optional public goods game along with the two usual strategies (viz. cooperate and defect), the frequencies of the strategies undergo global oscillations.
In contrast, we use a single (in effect, symmetric) diffusion coefficient---with the agents in the subpopulations playing two-player games (LG or PD)---to bring about the coexistence of the cooperators and the defectors over the entire population homogeneously. Regardless of the differences, it would be interesting to incorporate multi-player games like the public goods games and also multi-strategy games~\cite{2021_Sayantan_rspa} within  subpopulations in our model.

Our model is technically a multilayer network~\cite{2014_B_PR,2015_WWSP_EPJB,2016_IBA_NC, 2016_RD_PRE} where inside each node the individuals have an undirected all-to-all connection being modelled by the one-shot games, and between two nodes there is directed connection fashioned by the migration. Moreover, since the edges between any two nodes are stochastic functions of time, one could consider our directed multilayer network as temporal. Consequently, the emergence of cooperation in it could be seen as an extension of similar enhancement of cooperation in temporal networks~\cite{2020_LZSCLWL_NC}. A possible extension of our work could be to investigate how different multilayered network structures affect the emergence of cooperation in multiplayer games with many strategies. 

The random rewiring scheme used in the CML could remind the readers about similar works done with the small-world networks. Previous works on small-world networks, with agents placed on the nodes and playing the PD game, focus on the role of topology~\cite{2001_AK_PRE,2003_MA_PLA}, introduction of a new strategy~\cite{2005_WXCW_PRE}, and aspiration level of agents~\cite{2010YWW} in promoting cooperation in the network. In contrast,  we focus on building the simplest possible model to showcase the role of non-convergent dynamics and synchronization in bringing about cooperation in the entire network. Another key difference is that we have placed subpopulations (and not individual agents) on the nodes of the network. Placing individual agents on nodes of a small world network yields a lower fraction of cooperators for a higher rewiring probability~\cite{2001_AK_PRE,2003_MA_PLA}. This is at odds with our results, where the system attains full cooperation beyond the critical coupling strength regardless of the rewiring probability when the power law exponent $\alpha$ is less than 1; and for $\alpha=1$, one can have stable coexistence of cooperation and defectors, and the stability depends on rewiring probability $p$.

We emphasize once again that the replicator map used in this paper does not consider the network structure within the subpopulation, i.e., it is assumed that an agent interacts with all other agents within a subpopulation. Interestingly, a formalism of replicator equation on graphs has been proposed~\cite{Ohtsuki_2006} where the underlying structure of the population is taken into consideration. It would be interesting to extend our model by considering structured subpopulations in line with this formalism. Also, we realize that the replicator map being a deterministic equation, should be seen as the mean-field limit of a microscopic stochastic process~\cite{Mukhopadhyay2021}; in a finite population, stochastic effects  must have impact on the evolution of cooperation and its evolutionary stability~\cite{Nowak2004, Wu2013}. A stochastic differential equation approach~\cite{2014_constable_JTB} was adopted to deal with the finiteness of the deme size where the direction and the strength of selection varied from deme to deme. Naturally, in future, one may envisage investigating the stochastic phenomena, like fixation probability, in the CML with finite subpopulations.

Besides the pairwise interaction, multiplayer interactions where more than two individuals are involved is ubiquitous and very relevant in the study of the evolution of cooperation~\cite{Joshi1987, Boyd1988, 2010_Gokhale, Han2012, Gokhale2014}. The multiplayer interactions can not be reduced to the sum of pairwise interactions that makes it quite fascinating~\cite{Wu2013, Perc2013}. Naturally, non-linear fitness gets introduced through multiplayer interactions. Several studies on the effect of interacting group size have been performed and they highlight the detrimental impact on the evolution of cooperation~\cite{Joshi1987, Boyd1988}. Realistic interactions are spatially structured and are conveniently modelled by means of networks without all-to-all interactions; as could be guessed, their topologies have inevitable effects on the corresponding results. On a square lattice, enhanced spatial reciprocity can additionally promote cooperation in large groups~\cite{Szolnoki2011}. Our study can be extended for the multiplayer interactions by introducing the multiplayer replicator maps using public goods game or multiplayer PD~\cite{Kurokawa2010, Kurokawa2018, 2009_WNH_PNAS, 2012_Wang_JStat}. The $d$-player $n$-strategy interaction gives rise to $(d-1)^{n-1}$ number of internal equilibria at maximum for continuous replicator dynamics~\cite{2010_Gokhale, Gokhale2014}. The distribution and the number of stable equilibria was investigated for $d$-player two-strategy game with random payoff matrix; it was found that the number of stable equilibria is $\sqrt{d-1}$ asymptotically~\cite{Duong2016}. It is easy to see that the discrete replicator map can have a maximum of $(d-1)^{n-1}$ internal equilibria if it is extended for $d$-player $n$-strategy game. Thus, there is a possibility that with only two strategies, multiplayer PD can give rise to internal equilibria which imply the coexistence of cooperation and defection, and this could trigger even higher degree of cooperation in the CML with LG in some demes.

Before we conclude, we would also like to point out that some variants of the coupled replicator equations~\cite{2003_SC_PRE, 2004TRS, 2009G, 2013KG, 2014JL} model the dynamics of collective learning in a group of agents who may be faced with the social dilemmas~\cite{2002_MF_PNAS}. They show quasiperiodicity, limit cycles, intermittency, and chaos, among other rich dynamical features. In this setting, one should be able to explore collective phenomena like synchronization onto an attractor. Furthermore, one could also ponder upon how bounded rationality~\cite{1955_S_QJE,1998_Rubenstein, 2000_S_GEB}, mutation~\cite{2015_TS_PRE} and delay~\cite{2020_MMC_PRE}, and hypergames~\cite{2018_JCHL_PRE} in a deme would modify the results of this paper.

\bibliographystyle{apsrev4-2}
\bibliography{game_theory_shubhadeep}
\end{document}